\documentclass[12pt]{article}
 \usepackage[cp1251]{inputenc} 
 \usepackage[english, russian]{babel} 
 \usepackage{amsmath, latexsym, amssymb, bm, array, graphics, eucal,
 amsfonts,}
 \makeatletter

\renewcommand{\Re}{\mathop{\rm Re}}
\renewcommand{\Im}{\mathop{\rm Im\,}}
\renewcommand{\div}{\mathop{\rm div}}
 \makeatother
 
\usepackage[dvips]{graphicx}

\begin{document}
\thispagestyle{empty}
\large
\renewcommand{\abstractname}{Abstract}
\renewcommand{\refname}{\begin{center}
 REFERENCES\end{center}}
\newcommand{\mc}[1]{\mathcal{#1}}
\newcommand{\E}{\mc{E}}
\makeatother

\begin{center}
\bf MAGNETIC SUSCEPTIBILITY AND LANDAU DIAMAGNETISM OF
QUANTUM COLLISIONAL DEGENERATE PLASMAS
\end{center}
\begin{center}
  \bf A. V. Latyshev\footnote{$avlatyshev@mail.ru$} and
  A. A. Yushkanov\footnote{$yushkanov@inbox.ru$}
\end{center}\medskip

\begin{center}
{\it Faculty of Physics and Mathematics,\\ Moscow State Regional
University, 105005,\\ Moscow, Radio str., 10--A}
\end{center}\medskip

\begin{abstract}
With the use of correct expression of the electric
conductivity of quantum collisional degenerate plasmas
the kinetic description of a magnetic susceptibility is obtained and
the formula for calculation of Landau diamagnetism is deduced.

\medskip

electron gas
\end{abstract}

\begin{center}
\bf  Introduction
\end{center}

Magnetisation of electron gas in a weak magnetic fields
compounds of two independent parts (see, for example, \cite {Landau5}):
from the paramagnetic magne\-ti\-sa\-tion connected
with own (spin) magnetic
momentum of elect\-rons ({\it Pauli's para\-mag\-netism}, W. Pauli, 1927)
and from the diamagnetic mag\-ne\-ti\-sation connected with
quantization of
orbital movement of elect\-rons in a magnetic field ({\it
Landau diamagnetism}, L. D. Landau, 1930).

Landau diamagnetism  was considered till now for a gas of the free
elect\-rons. It has been thus shown, that together with original
approach develo\-ped by Landau, expression for diamagnetism of electron
gas can be obtained on the basis of the kinetic approach
\cite {Silin}.

The kinetic method gives opportunity to calculate the trans\-verse
die\-lect\-ric permeability.  On the basis of this quantity its possible
to obtain
the value of the diamagnetic response.

However such calculations till now
were carried out only for collisional\-less case. The matter is that
correct expression for the transverse dielectric
permeability of quantum plasma existed till
 now only for gas of the free
electrons. Expression known till now for the transverse dielectric
perme\-abi\-lity in  a collisional case gave incorrect transition to
the classical case \cite {Kliewer}. So this expression  were accordingly
incorrect.

Central result from \cite{Datta} connects the mean orbital
magnetic moment, a thermodynamic property, with the electrical
resistivity, which characterizes transport properties of
material. In this work
was discussed the important problem of  dissipation (collisions)
influence on  Landau diamagnetism. The analysis of this problem
is given with
use of exact expression of transverse conductivity of quantum plasma.

In work \cite{Kumar} is shown that a classical system of charged
particles moving on a finite but unbounded surface (of a sphere)
has a nonzero orbital diamagnetic moment which can be large.
Here is considered a non-degenerate system with the degeneracy
temperarure much smaller than the room temperature, as in the
case of a doped high-mobility semiconductor.

In work \cite{LY}  for the first time the expression   for
the quantum transverse dielectric
permeability of collisional degenerate plasma has been derived. The
obtained in \cite {LY} expression for
trans\-ver\-se dielectric permeability satisfies
to the necessary requirements of com\-pa\-ti\-bility.

In the present work for the first time  with use of correct
expression for the transverse conductivity \cite {LY} the
kinetic description of a magnetic susceptibility of quantum
collisional degenerate plasmas is given. The formula for
calculation of Landau
diamagnetism for degenerate collisional plasmas is deduced.

\begin{center}
  \bf 2. Magnetic susceptibility of quantum degenerate plasmas
\end{center}

Magnetization vector $\mathbf{M}$ of electron plasma
is connected with current density $\mathbf{j}$ by the following
expression \cite {Landau8}
$$
{\bf j}=c\, {\rm rot}\,\mathbf{M},
$$
where $c$ is the light velocity.

Magnetization vector $\mathbf{M}$ and a magnetic field
strength
$\mathbf{H}=\rm rot \mathbf{A}$ are connected by the expression
$$
{\bf M}=\chi\,{\bf H}=\chi\,{\rm rot}\,{\bf A},
$$
where $\chi$ is the magnetic susceptibility, $\mathbf{A}$ is the
vector potential.

From these two equalities for current density we have
$$
\mathbf{j}={c}\, {\rm rot}\,\mathbf{M}=
c\,\chi {\rm rot}\, \big({\rm rot}\,\mathbf{A}\big)=
c\,\chi\, \big[{\bf \nabla}({\bf \nabla}\cdot{\bf A})-
{\mathbf{\triangle}}\mathbf{A}\big].
$$

Here $\Delta$ is the Laplace operator.

Let the scalar potential is equal to zero.
Vector potential we take ortho\-gonal
to the direction of a wave vector $\mathbf{q}$
($\mathbf{q}\mathbf{A}=0$) in the form of a  harmonic wave
$$
\mathbf{A}(\mathbf{r},t)=\mathbf{A}_0
e^{i(\mathbf{q} \mathbf{r}-\omega t)}.
$$

Such vector field is solenoidal
$$
\div \mathbf{A}=\nabla\mathbf{A}=0.
$$

Hence, for current density we receive equality
$$
{\bf j}=-c\,\chi \Delta \mathbf{A}=c\,\chi\,q^2 \mathbf{A}.
$$

On the other hand, connection of electric field  $\mathbf{E}$ and vector
potential $\mathbf{A}$ is as follows
$$
\mathbf{E}=-\dfrac{1}{c}\dfrac{\partial \mathbf{A}}{\partial t}=
\dfrac{i\omega}{c}\mathbf{A}.
$$
It is leads to the relation
$$
\mathbf{j}=\sigma_{tr}\mathbf{E}=\sigma_{tr}\dfrac{i\omega}{c}
\mathbf{A},
$$
where $\sigma_{tr}$ is the transverse electric conductivity.

For our case from (1.1) and (1.2) we obtain
following expression for the magnetic susceptibility
$$
\chi=\dfrac{i\omega}{c^2 q^2}\sigma_{tr}.
\eqno{(1.1)}
$$

Expression of transversal conductivity of degenerate collisional
plasmas it is defined by the general formula \cite{LY}:
$$
\sigma_{tr}({\bf q},\omega,\nu)=\sigma_0\dfrac{i \nu}{\omega}\Big(1+
\dfrac{\omega J({\bf q},\omega,\nu)+i \nu J({\bf q},0,0)}
{\omega+i \nu}\Big),
\eqno{(1.2)}
$$
where $\sigma_0$ is the static conductivity,
$\sigma_0={e^2N}/{m\nu}$, $N$ is the concentration (number density)
of plasmas particles, $e$ and $m$ is the electron charge and mass,
$\nu$ is the effective collisional frequency of plasmas particles,
$$
J({\bf q},0,0)=\dfrac{\hbar^2}{8\pi^3mN}\int
\dfrac{f_{\bf k}-f_{\bf k-q}}
{\E_{\bf k}-\E_{\bf k-q}}{\bf k}_\perp^2d^3k,
$$

$$
J({\bf q},\omega,\nu)=\dfrac{\hbar^2}{8\pi^3mN}
\int \dfrac{f_{\bf k}-f_{\bf k-q}}
{\E_{\bf k}-\E_{\bf k-q}-\hbar(\omega+i \nu)}{\bf k}_\perp^2d^3k,
$$
$$
f_{\bf k}=\Theta(\E_{\bf k}-\E_F),
$$
$\Theta(x)$ is the function
of Heaviside,
$\E_{\bf k}={\hbar^2{\bf k}^2}/{2m}$ is the electron energy,
$\E_F={mv_F^2}/{2}$ is the electron energy on Fermi surface,
$v_F$ is the electron velocity on Fermi surface, which
is considered spherical, $\hbar$ is the Planck's constant,
$$
\Theta(x)=\left\{\begin{array}{c}
                   1,\qquad x>0, \\
                   0, \qquad x<0,
                 \end{array}
\right.
$$

$$
{\bf k}_\perp^2={\bf k}^2-\Big(\dfrac{{\bf kq}}{q}\Big)^2.
$$

According to (1.1) and (1.2) magnetic susceptibility of the quantum
collisional degenerate plasmas it is equal
$$
\chi({\bf q},\omega,\nu)=-\dfrac{e^2N}{mc^2q^2}\Big(1+
\dfrac{\omega J({\bf q},\omega,\nu)+i \nu J({\bf q},0,0)}{\omega+i \nu}\Big).
\eqno{(1.3)}
$$

From the formula (1.3) it is visible, that at $\omega=0$ frequency
of collisions plasma particles $ \nu $ drops out of the formula (1.3).
Hence, the magnetic susceptibility in a static limit does not depend from
frequencies of collisions of plasma and the following form also has:

$$
\chi({\bf q},0,\nu)=-\dfrac{e^2N}{mc^2q^2}\Big[1+
\dfrac{\hbar^2}{8\pi^3mN}\int \dfrac{f_{\bf k}-f_{\bf k-q}}
{\E_{\bf k}-\E_{\bf k-q}}{\bf k}_\perp^2d^3k\Big].
\eqno{(1.4)}
$$

From expression (1.3) it is visible, that a magnetic susceptibility in
collisionless quantum degenerate plasma is equal:
$$
\chi({\bf q},\omega,0)=-\dfrac{e^2N}{mc^2q^2}\Big[1+
\dfrac{\hbar^2}{8\pi^3mN}\int \dfrac{f_{\bf k}-f_{\bf k-q}}
{\E_{\bf k}-\E_{\bf k-q}-\hbar\omega}{\bf k}_\perp^2d^3k\Big].
\eqno{(1.5)}
$$
At $ \omega\to  0$ the formula (1.5) passes in the formula (1.4).

Let's deduce the formula for calculation of a magnetic susceptibility
of quantum collisional degenerate plasmas.

After obvious linear replacement of variables the formula for integral
$J({\bf q}, \omega, \nu)$ will be transformed to the form
$$
J=\dfrac{\hbar^2}{8\pi^3mN}
\int \dfrac{(\E_{\bf k+q}+\E_{\bf k-q}-2\E_{\bf k})
f_{\bf k}{\bf k}_\perp^2d^3k}
{[\E_{\bf k}-\E_{\bf k-q}-\hbar(\omega+i \nu)]
[\E_{\bf k+q}-\E_{\bf k}-\hbar(\omega+i \nu)]}.
\eqno{(1.6)}
$$

Let's enter dimensionless variables
$$
z=\dfrac{\omega+i \nu}{k_Fv_F}=x+iy, \quad
x=\dfrac{\omega}{k_Fv_F}, \quad y=\dfrac{\nu}{k_Fv_F}, \quad
Q=\dfrac{q}{k_F}.
$$
Then
$$
\E_{\bf k}-\E_{\bf k-q}-\hbar(\omega+i \nu)=2\E_FQ\Big(K_x-\dfrac{z}{Q}-
\dfrac{Q}{2}\Big),
$$
$$
\E_{\bf k+q}-\E_{\bf k}-\hbar(\omega+i \nu)=2\E_FQ\Big(K_x-\dfrac{z}{Q}+
\dfrac{Q}{2}\Big),
$$
$$
\E_{\bf k+q}+\E_{\bf k-q}-2\E_{\bf k}=2\E_FQ^2.
$$
Considering, that for degenerate plasmas $k_F^3=3\pi^2N $, on
the basis (1.6) it is received
$$
J(Q,z)=\dfrac{3}{8\pi}\int \dfrac{f_{\bf K}{\bf K}_\perp^2 d^3K}
{(K_x-z/Q)^2-(Q/2)^2},
\eqno{(1.7)}
$$
where
$$
f_{\bf K}=\Theta(1-{\bf K}^2),\qquad {\bf K}_\perp^2=K_y^2+K_z^2.
$$

Now the formula (1.3) for a magnetic susceptibility  can be written down
in the form
$$
\chi(Q,x,y)=-\dfrac{e^2v_F}{3\pi^2\hbar c^2Q^2}\Big(1+
\dfrac{xJ(Q,z)+iyJ(Q,0)}{x+iy}\Big).
\eqno{(1.8)}
$$

Here according to (1.7)
$$
J(Q,z)=\dfrac{3}{16}\int\limits_{-1}^{1}\dfrac{(1-\tau^2)^2d\tau}
{(\tau-z/Q)^2-(Q/2)^2}=-\dfrac{5}{8}+\dfrac{3z^2}{4Q^2}+\dfrac{3Q^2}{32}+
$$
$$
+\dfrac{3}{16Q}\Big[\Big(1-\dfrac{z^2}{Q^2}\Big)^2+\dfrac{Q^4}{16}-
\dfrac{Q^2}{2}+\dfrac{3z^2}{2}\Big]\ln\dfrac{(1-Q/2)^2-(z/Q)^2}
{(1+Q/2)^2-(z/Q)^2}-
$$
$$
-\dfrac{3zQ}{32}\Big[1+\dfrac{4}{Q^2}\Big(1-\dfrac{z^2}{Q^2}\Big)\Big]
\ln\dfrac{(1-z/Q)^2-(Q/2)^2}{(1+z/Q)^2-(Q/2)^2},
$$
$$
J(Q,0)=-\dfrac{5}{8}+\dfrac{3Q^2}{32}+\dfrac{3(4-Q^2)^2}{256Q}
\ln\Big(\dfrac{2-Q}{2+Q}\Big)^2.
$$

\begin{center}
\bf 2. Landau diamagnetism of quantum degenerate collisionless
plasmas
\end{center}

Landau diamagnetism  in collisionless  plasma is usually
 defined as a magnetic susceptibility in a static limit
for a homogeneous external magnetic field.
Thus the diamagnetism value can be found by means of (1.1)
through two  non-commutative limits
$$
\chi_L=\lim\limits_{q\to 0}\Big[\lim\limits_{\omega\to 0}^{}
\chi(q,\omega,\nu=0)\Big].
\eqno{(2.1)}
$$

Into collisionless  plasma
this expression (2.1) should lead to the known formula of
Landau's diamagnetism
$$
\chi_L=-\dfrac{1}{3}
\left(\dfrac{e\hbar}{2m c}\right)^2\dfrac{p_Fm}{\pi^2\hbar^3}=
-\dfrac{e^2v_F}{12\pi^2\hbar c^2}.
\eqno{(2.2)}
$$

Let's deduce the formula of Landau's diamagnetism (2.2) by means
of expression (1.8). At $z = + iy=0$ from the formula (1.8)
for magnetic susceptibility of the quantum
collisionless degenerate plasmas we receive the following expression
$$
\chi(Q)=-\dfrac{e^2v_F}{3\pi^2\hbar c^2Q^2}\Big[1+
\dfrac{3}{8\pi}\int
\dfrac{f_{\bf K}{\bf K}_\perp^2d^3K}{K_x^2-(Q/2)^2}\Big]=
$$
$$
=-\dfrac{e^2v_F}{3\pi^2\hbar c^2Q^2}\Big[1+
\dfrac{3}{16}\int\limits_{-1}^{1}
\dfrac{(1-\tau^2)^2d\tau}{\tau^2-(Q/2)^2}\Big],
$$
or, in explicit form
$$
\chi(Q)=-\dfrac{e^2N}{mc^2k_F^2Q^2}\Big[\dfrac{3}{8}+
\dfrac{3}{32}Q^2+\dfrac{3(Q^2-4)^2}{128Q}\ln\Big|\dfrac{2-Q}{2+Q}\Big|\Big].
\eqno{(2.3)}
$$
Noticing, that at small $Q $
$$
\dfrac{1}{Q}\ln\dfrac{1-Q/2}{1+Q/2}=-1-\dfrac{Q^2}{12}-\cdots,
$$
on the basis of (2.3) we found the known expression for diamagnetism
of Landau for degenerate electronic gas
$$
\chi_L=\lim\limits_{Q\to 0}\chi(Q)=-\dfrac{e^2k_F}{12\pi^2mc^2}=
-\dfrac{e^2v_F}{12\pi^2c^2\hbar}.
\eqno{(2.4)}
$$

\begin{figure}[h]
\begin{flushleft}
\includegraphics[width=15.0cm, height=9cm]{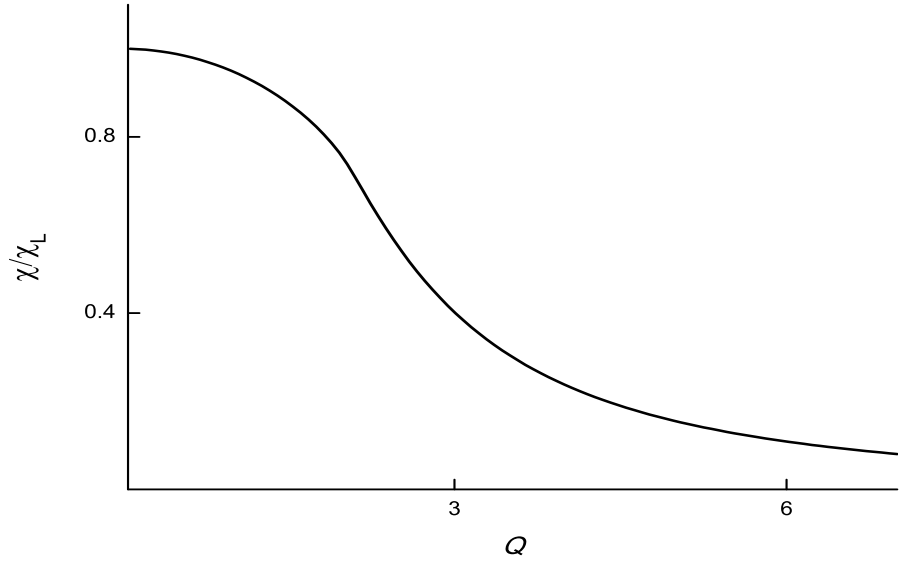}
\caption{Diamagnetic susceptibility for the case $\omega=0$ into
collisionless plasmas ($\nu=0$).
}\label{rateIII}
\end{flushleft}
\end{figure}

Having divided (2.3) on (2.4), we receive (see fig. 1) the relative
magnetic susceptibility for quantum collisionless plasmas in
static limit
$$
\dfrac{\chi(Q)}{\chi_L}=\dfrac{4}{Q^2}
\Big[\dfrac{3}{8}+
\dfrac{3}{32}Q^2+\dfrac{3(Q^2-4)^2}{128Q}\ln\Big|\dfrac{2-Q}{2+Q}\Big|\Big].
$$

\begin{center}\bf
  3. The analysis of results
\end{center}

Let's present the formula (1.8) in the form
$$
\dfrac{\chi(Q,z)}{\chi_L}=\dfrac{4}{Q^2}\Big(1+
\dfrac{xJ(Q,z)+iyJ(Q,0)}{x+iy}\Big).
\eqno{(3.1)}
$$

For graphic research of a magnetic susceptibility we will be
to use the formula (3.1).

From fig. 1 it is obvious, that in quantum collisionless plasma
($ \nu=0$) in a static limit ($ \omega=0$) the magnetic
susceptibility is function of wave number monotonously decreasing to zero.

On fig. 2 and 3 the dependence of a magnetic susceptibility  of
collisionless plasmas as function of wave number (fig. 2)
or function of dimensionless frequency of oscillations of an
electromagnetic field $x$ (fig. 3) is presented.

From fig. 2 it is clear, that the magnetic susceptibility is
monotonously decreasing function of wave number at all values of frequency
of oscillations of electromagnetic field. Thus for all $x <1$
($ \omega <\omega_p $) values of a magnetic susceptibility that
more than the quantity of frequency of oscillations of the
electromagnetic field is mores.

From fig. 3 it is clear, that a magnetic susceptibility as function
of frequencies of oscillations of a field has
the maximum near to frequency $ \omega=Q\omega_p $
and with growth $Q $ moves to the right.

On fig. 4 and 5 the dependence of real (fig. 4) and imaginary (fig. 5) parts of
the magnetic susceptibility from the dimensionless frequency of
oscillations of the field in the case $Q=0.5$ are presented.

From fig. 4 it is clear, that the real part has a maximum,
which is
\begin{figure}[h]
\begin{flushleft}
\includegraphics[width=15.0cm, height=10cm]{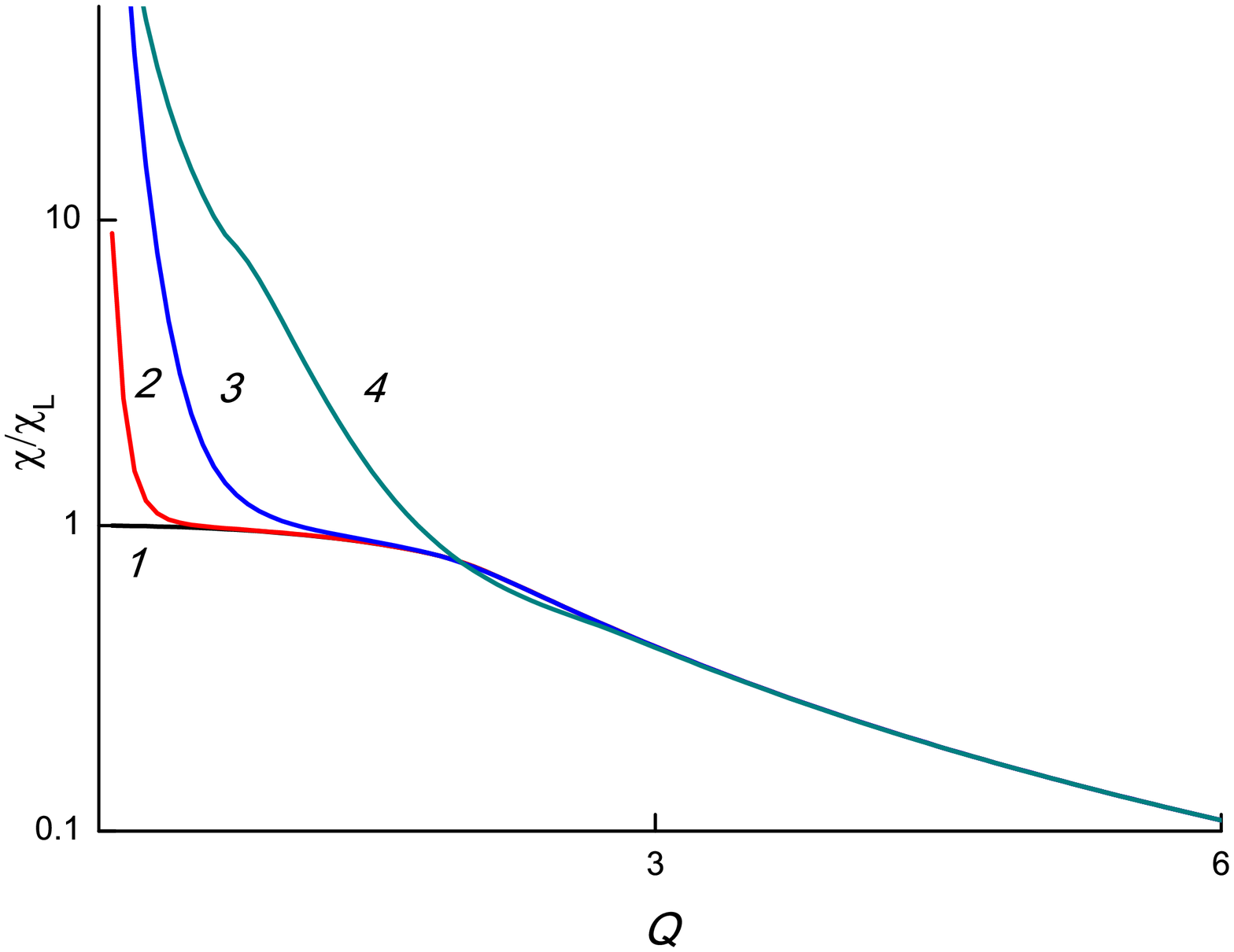}
\caption{Diamagnetic susceptibility of collisionless  plasmas,
curves $1$,$2$,$3$ and $4$ correspond to paramater quantities
$x=0, 0.01, 0.1$ and $x=1$.
}\label{rateIII}
\end{flushleft}
\begin{flushleft}
\includegraphics[width=15.0cm, height=10cm]{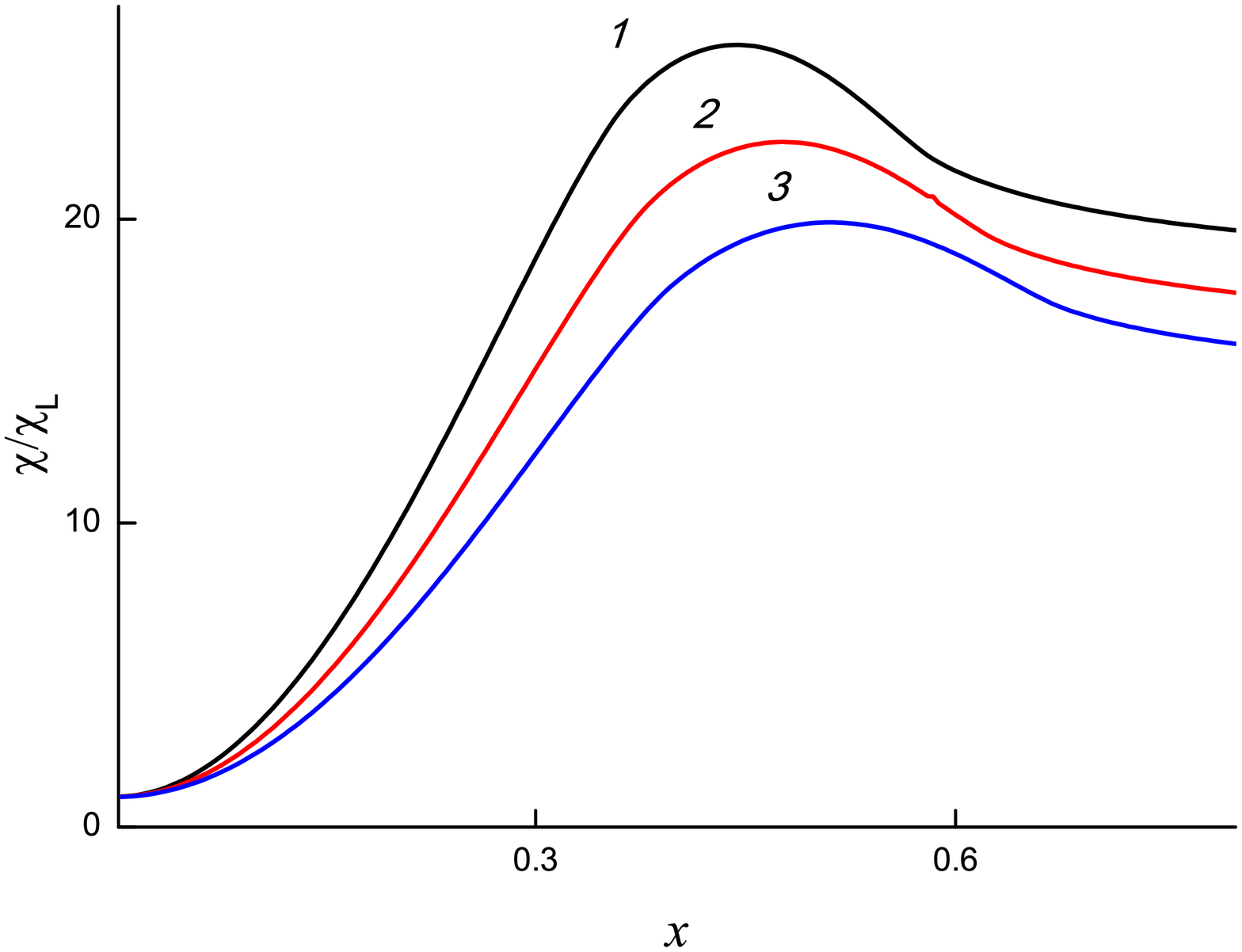}
\caption{Diamagnetic susceptibility of collisionless  plasmas,
curves $1$, $2$, $3$ correspond to paramater quantities
$Q=0.45, 0.50, 0.55$.
}\label{rateIII}
\end{flushleft}
\end{figure}
\clearpage
\noindent displaced to the right with growth of frequency of
collisions of plasmas.

Independently of the frequency of collisions of plasmas particles with
growth of frequency of oscillations of electromagnetic field quantity
of the real part of the magnetic susceptibility leaves from above on
the asymptotic
$$
\lim\limits _ {x\to 0} \Re\Big (\dfrac {\chi (Q, x, y)} {\chi_L} \Big) =
\dfrac {4} {Q^2}.
$$

Not resulting necessary graphics we will inform, that with reduction
quantity of wave number the maximum of  magnetic susceptibility
moves to the left and becomes sharp at small values of frequency
collisions of plasmas particles. With growth of frequency of collisions
the maximum starts to smooth out and vanishes.

From fig. 5 it is obvious, that the imaginary part of  magnetic
sus\-cep\-tibility as function of dimensionless frequency of
oscillations of electro\-magne\-tic field has a minimum.
This minimum moves to the left with growth
collisions frequency of plasmas particles.
With the growth of the dimensionless
frequencies of oscillations of electromagnetic field an imaginary part
of the magnetic susceptibilities leaves from below on the asimptotyc
$ \Im (\chi/\chi_L) =0$. We will notice, that a minimum of an imaginary part not
vanishes with growth as frequencies of collisions of particles of plasma, and
of dimensionless wave number.

Let's notice, that the frequency of collisions of
plasma particles there is less, the
more values of the real and imaginary parts magnetic susceptibilities
turn out.

\begin{figure}[h]
\begin{flushleft}
\includegraphics[width=15.0cm, height=10cm]{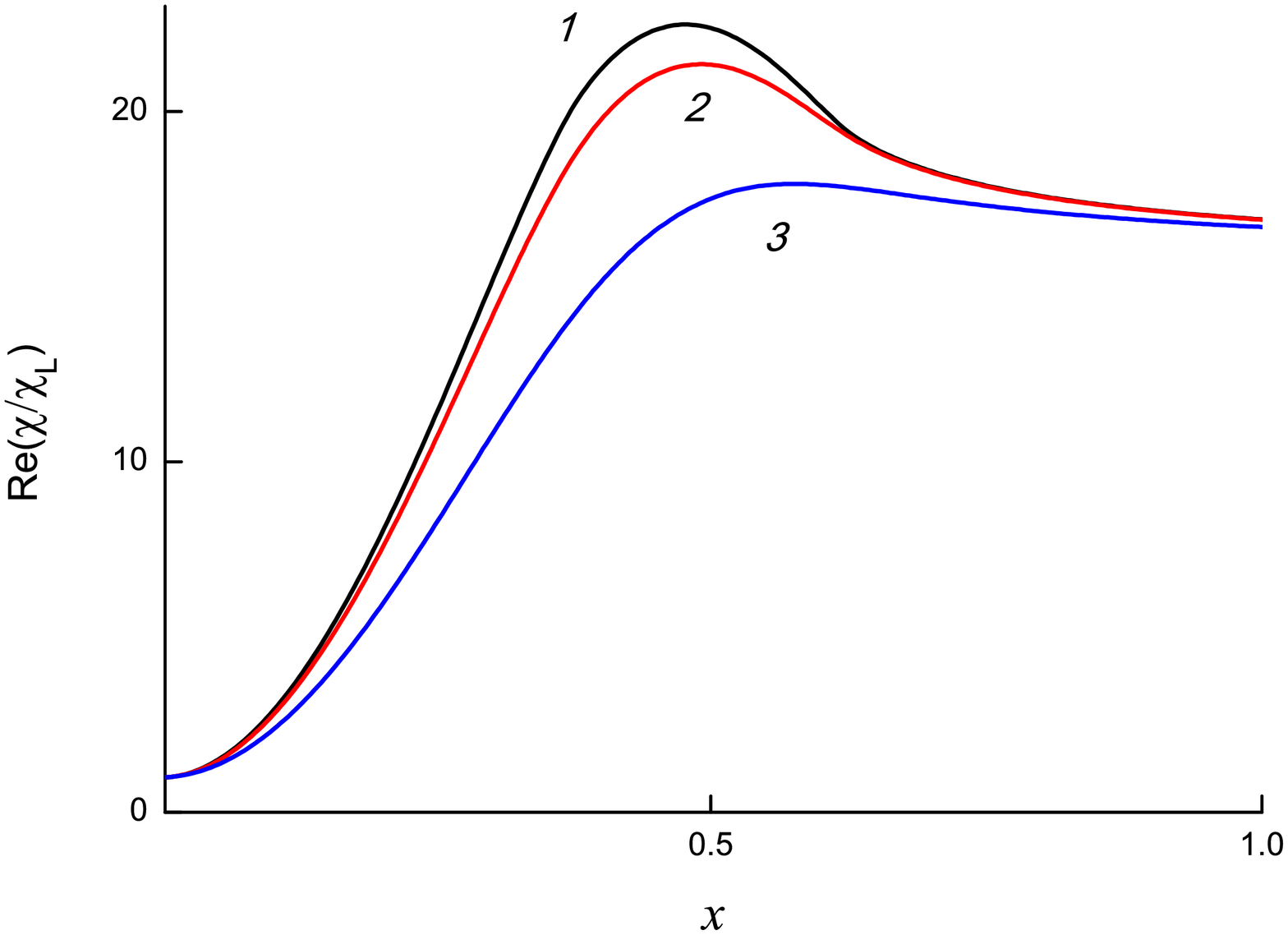}
\caption{Real part of diamagnetic susceptibility for case
$Q=0.5$, curves $1,2,3$ correspond to parameter values
$y=10^{-3}, 10^{-2}, 10^{-1}$.
}\label{rateIII}
\end{flushleft}
\begin{flushleft}
\includegraphics[width=15.0cm, height=10cm]{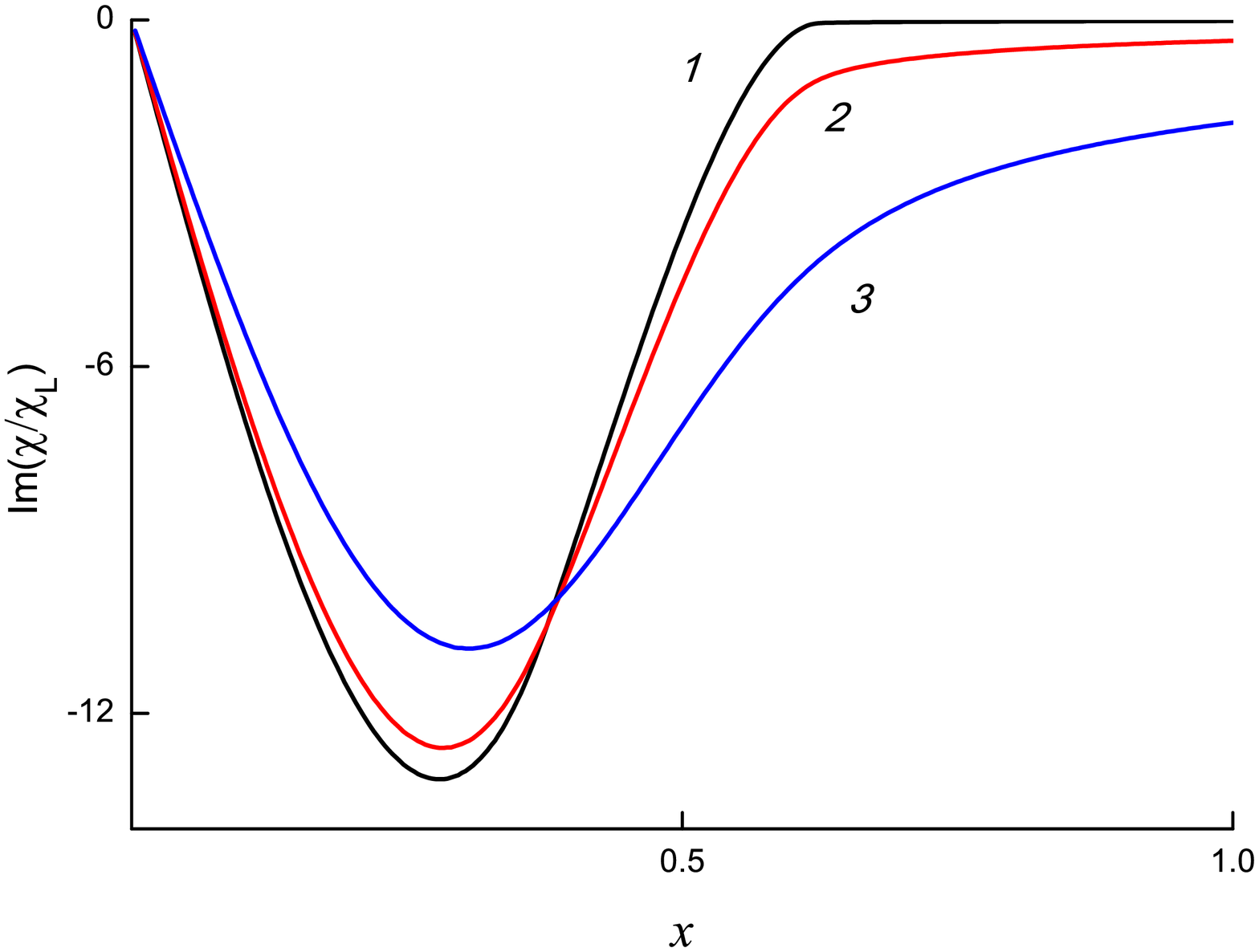}
\caption{Imaginary part of diamagnetic susceptibility for case
$Q=0.5$, curves $1,2,3$ correspond to parameter values
$y=10^{-3}, 10^{-2}, 10^{-1}$.
}\label{rateIII}
\end{flushleft}
\end{figure}
\clearpage

\begin{center}
  \bf 4. Conclusions
\end{center}

In the present work the kinetic description of magnetic
susceptibilities of quantum collisional degenerate
plasmas with use before deduced correct formulas for
electric conductivity of quantum plasma is given.

Influence of the collisions of plasma particles on the magnetic
susceptibility is found out.
Thereby the answer to a question on influence is given to the
question of
dissipation on Landau diamagnetism  put in work \cite {Datta}.
For collisionless plasmas with the help
the kinetic approach the known formula of Landau diamagnetism is
deduced.

\end{document}